\documentclass[review,3p]{elsarticle}

\usepackage{amssymb}
\usepackage{amsmath}
\usepackage{amsthm}
\usepackage{amsfonts}
\usepackage{bm}
\usepackage{subcaption}
\usepackage{ulem}
\usepackage[figurename=Fig.]{caption}

\usepackage{multicol}
\usepackage{color}
\usepackage{xspace}
\usepackage{pdfwidgets}
\usepackage{enumerate}
\newcommand{\datrevise}[1]{\textcolor{black}{#1}}

\newcommand{\diluka}[1]{\textcolor{black}{#1}}

\makeatletter
\def\bs{\expandafter\@gobble\string\\}
\def\lb{\expandafter\@gobble\string\{}
\def\rb{\expandafter\@gobble\string\}}
\def\@pdfauthor{C.V.Radhakrishnan}
\def\@pdftitle{elsarticle.cls -- A documentation}
\def\@pdfsubject{Document formatting with elsarticle.cls}
\def\@pdfkeywords{LaTeX, Elsevier Ltd, document class}

\begin{document}

\begin{frontmatter}



\title{Multi-object Tracking with an Adaptive Generalized Labeled Multi-Bernoulli Filter\tnoteref{t1}}

\tnotetext[t1]{This work was partly supported by the Joint-scholarship between Ministry of Education and Training Vietnam and Curtin International Postgraduate Research Scholarship (MOET-CIPRS), and the Ministry of Science and ICT (MSIT), Korea, under the ICT Creative Consilience program (IITP-2021-2020-0-01819) supervised by the Institute for Information \& communications Technology Planning \& Evaluation  (IITP).}


\author[1]{Cong-Thanh Do\corref{cor1}
}
\ead{thanh.docong@postgrad.curtin.edu.au }
\cortext[cor1]{Corresponding author}

\author[1]{Tran Thien Dat Nguyen}
\ead{t.nguyen172@postgrad.curtin.edu.au }

\author[1]{Diluka Moratuwage}
\ead{diluka.moratuwage@curtin.edu.au}

\author[1]{Changbeom Shim}
\ead{changbeom.shim@curtin.edu.au}

\author[2]{Yon Dohn Chung}
\ead{ydchung@korea.ac.kr}

\address[1]{School of Electrical Engineering, Computing, and Mathematical Sciences,
Curtin University, Bentley, WA 6102, Australia}

\address[2]{Department of Computer Science and Engineering, Korea University, Seoul, 02841, Republic of Korea}

\begin{abstract}

\datrevise{The challenges in multi-object tracking mainly} \diluka{stem from the random variations in the cardinality and states of objects during the tracking process. Further, the information on locations where the objects appear, their detection probabilities, and the statistics of the sensor's false alarms significantly influence the tracking accuracy of the filter.} \datrevise{However, this information is usually assumed to be known and provided by the users. In this paper, we propose an adaptive generalized labeled multi-Bernoulli (GLMB) filter which can track multiple objects without prior knowledge of the aforementioned information. Experimental results show that the performance of the proposed filter is comparable to an ideal GLMB filter supplied with correct information of the tracking scenarios.}
\end{abstract}


\end{frontmatter}


\section{Introduction}
Multi-object tracking is the problem of estimating the number of objects and their states from noisy measurements. The quality of estimation not only depends on the approximation of the multi-object dynamic model but also on the data association uncertainty, false alarms and miss-detections. In Bayesian paradigm, there are three main approaches to address the multi-object tracking problem: the joint probabilistic data association (JPDA) \cite{Fortmann1983Sonar}, the multi-hypothesis tracking (MHT) \cite{Reid1979AnAlgorithm}, and the random finite set (RFS) \cite{Mahler2007Statistical,Mahler2014Advances}. 

In RFS approach, the multi-object tracking problem is tackled from a top-down manner with the multi-object dynamics (e.g., object motions, births, deaths and spawning) and measurement statistics  are jointly taken into account. Specifically, in this approach the multi-object state and measurements are represented by sets. The set of multiple objects is considered as a random variable and its probability density is propagated through time using Bayes recursion. Nevertheless, due to the complexity of the multi-object density, approximations or only some of its statistics (e.g., first moment and cardinality distribution) are propagated for tractability.

In RFS without labels, trajectories cannot be estimated in principle manner since without object identities, one cannot associate object states in the temporal domain to form trajectories without using heuristics. The recently proposed labeled RFS, which is essentially a marked RFS with distinct marks \cite{Vo2013Labeled}, can effectively resolve this issue. Further, it also allows top-down estimation of objects lineage  since ancestral information can be embedded into the object labels. Among labeled RFS filters, generalized labeled multi-Bernoulli (GLMB) filter \citep{Vo2017AnEfficient,Vo2014Labeled} is the most popular in the modern literature, which is also proven to be a Bayes optimal multi-object tracking filter \cite{Mahler2019Exact}.

In multi-object tracking, in addition to kinematic observation noise, the measurement set is also corrupted by false alarms (clutter) and miss-detection. In standard practice, clutter is usually assumed to be a Poisson process characterized by a clutter rate parameter (clutter is assumed to be uniformly distributed in the measurement space). In most cases, detection probability of objects and clutter rate  are considered as temporally fixed and known a priori. However, this assumption is unrealistic in many practical applications (e.g., radar-based tracking) where detection probability and clutter rate are not time-constants. Supplying incorrect parameters may deteriorate the performance of the multi-object tracking algorithms significantly. Hence, in practice, these parameters are either estimated from training data or manually tuned \cite{Mahler2011CPHD}. However, training data is not always available and manual tuning can be a tedious task. To this extent, a robust cardinalized probability hypothesis density (CPHD) filter is developed to jointly estimate the detection probability and clutter rate along with object states \cite{Mahler2011CPHD}. Despite being computationally efficient (data association is not required), this filter does not include object identities. Therefore, it is impossible to estimate the object trajectories in a principled manner.

Further, it is common that the locations where objects appear are assumed to be known by the filters. However, this assumption may not hold in many cases where new objects could appear anywhere in the tracking region. To resolve this issue, authors of \cite{Reuter2014TheLabeled} proposed a method of using measurements from the last time step to initiate new tracks at the current time step. Nevertheless, this method requires information of the measurement-to-track association.


\datrevise{To efficiently handle unknown detection probability, clutter rate and birth locations within the labeled RFS filtering framework, in this paper we propose an online adaptive GLMB filter that:
\begin{itemize}
    \item does not require prior knowledge about the detection probability and clutter rate;
    \item can initiate new tracks without prior knowledge of object birth locations.
\end{itemize}}
Specifically, the detection probability and clutter rate are estimated by an independent robust CPHD filter \cite{Mahler2011CPHD} which are then bootstrapped to a standard GLMB filter \cite{Vo2017AnEfficient} for multi-object state estimation. The measurement-driven birth model in \cite{Reuter2014TheLabeled} is used to propose new tracks at each time step with the information on measurement-to-track association obtained from the GLMB filter. Furthermore, the proposed filter inherits all advantages of the GLMB filter.

\section{Background }\label{sec:Background}
\subsection{RFS multi-object tracking filters}
Theoretically, an RFS Bayesian multi-object tracking filter propagates a probability density function of the multi-object state through time using the Bayes recursion. However, due to the complexity of this density, several approximate filtering solutions have been proposed to maintain the tractability. In particular, the probability hypothesis density (PHD) filter \cite{Mahler2003Multitarget} only propagates the first-order statistical moment of the multi-object density. Conversely, the CPHD filter \cite{Mahler2007PHD} further improves the performance of the PHD filter by additionally propagating the cardinality distribution. Unlike PHD and CPHD filters, which are based on moment approximation, the MeMBer filter \cite{Mahler2007Statistical} propagates a tractable number of hypothesized tracks, where each track is characterized by an existence probability and a spatial distribution of its state. Nonetheless, the above multi-object filters do not include object identities, and therefore, their use is limited in applications that require object trajectory estimation.

Labeled RFS filters propagate the object identities within the filter, thereby implicitly estimating the object trajectories. The GLMB filter \cite{Vo2013Labeled,Vo2014Labeled,Vo2017AnEfficient} is the first provably Bayes-optimal \citep{Mahler2014Advances}, labeled RFS filter, that solves the multi-object filtering problem in exact closed-form \cite{Vo2013Labeled}. Recently, the GLMB filter has shown that it can track over a million objects \cite{Beard2020ASolution}. An approximation of the GLMB filter, the labeled multi-Bernoulli (LMB) filter, was proposed in \cite{Reuter2014TheLabeled} to improve computational efficiency further, with the expense of reduced tracking accuracy. The labeled RFS filters have been used in the literature to solve various practical problems in multi-object tracking, for example, in computer vision \cite{Rathnayake2020Online,Ong2020ABayesian,Kim2019ALabeled,Nguyen2018Online,Hadden2017Stem}, simultaneous localization and mapping (SLAM) \cite{Moratuwage2019Delta} in robotics, multi-sensor management \cite{Beard2017Void,Panicker2020Tracking} and multiple drones control \cite{Nguyen2019Online,Nguyen2020MultiObjective}. Further extensions of the GLMB filter have also been proposed for track-before-detect (TBD) \cite{Papi2015Generalized}, spawning of objects \cite{Nguyen2021Tracking,Bryant2018AGeneralized}, merged measurements \cite{Beard2015Bayesian}, extended objects \cite{Beard2016Multiple}, multi-sensor tracking \cite{Vo2019MultiSensor}, and multi-object smoothing \cite{Vo2019AMultiScan}.

\subsection{Adaptive multi-object tracking}
\diluka{In this work, adaptive multi-object tracking is defined as  tracking multiple objects in an environment using noisy measurements, where the detection probability, clutter profile and the birth locations in the field of view are unknown. In general, this is a complex problem to solve due to the increased state space dimension and the uncertainty involved. Current approaches estimate the unknown probability of detection by augmenting it to the single-object state and jointly estimating them both. On the other hand, the clutter statistics are estimated by considering measurement clutter as a different type of object, whose behaviour is independent of the actual objects. The statistics of clutter are then inferred using the estimated set of clutter objects.
}


\datrevise{Several methods have been proposed \diluka{in the literature} for the RFS-based filters to tackle \diluka{the adaptive} \diluka{multi-object} tracking problem. In \cite{Mahler2010CPHD}, a variation of CPHD filter was proposed to accommodate the unknown probability of detection, which is then further extended to additionally estimate the clutter rate in \cite{Mahler2011CPHD}. The \diluka{latter} is referred to as the robust CPHD filter. A method in \cite{Beard2013Multitarget} uses the robust CPHD filter to estimate the clutter rate which is then bootstrapped into another standard CPHD filter to estimate the multi-object state. Several \diluka{other} solutions have also been proposed to estimate \diluka{the clutter rate} and detection probability online based on other filters including Kronecker delta mixture and Poisson (KDMP) and MeMBer filters \cite{Correa2016Estimating,Vo2013Robust,Kim2018Visual}. \diluka{More recently,} a GLMB filter which can estimate object detection probability and clutter profile has been proposed in \cite{Punchihewa2018Multiple}. \diluka{Nonetheless, since data association needs to be performed for both actual and measurement clutter objects, this filter is relatively expensive.}}



\diluka{On the other hand, to resolve unknown birth locations, Beard et al. proposed a partially uniform birth model in \cite{Beard2013APartial} that accommodates the measurement origins (in the previous time step) in the state space to initiate birth objects for the PHD and CPHD filters.}
\diluka{In \cite{Ristic2012Adaptive}, an adaptively varying birth intensity model (at each scan) has been proposed by distinguishing the persistent and newborn objects.} \diluka{Subsequently, Reuter et al. proposed measurement-driven birth models for the cardinality balanced MeMBer filter in \cite{Reuter2013Cardinality} and the LMB filter in \cite{Reuter2014TheLabeled}. Recently, a tractable measurement-driven birth model has been proposed for the multi-sensor GLMB filter \cite{Trezza2021Multisensor}.}

\section{Multi-object dynamic and measurement models} \label{sec:The-proposed-filter} 
\diluka{This section introduces some preliminaries on RFSs, multi-object Bayes filter, labeled multi-object transition and measurement models used in the GLMB filter.}
\subsection{Notations}

To facilitate our discussions, we adopt the same notations scheme used in \cite{Vo2013Labeled}. Specifically, the single-object states are represented by lower case letters, i.e. $x$ and $\bm{x}$, while multi-object states are represented by upper case letters, i.e. $X$ and $\bm{X}$. Note that the bold letters are used for labeled states. The single-object state space, label space and measurement space are respectively denoted as $\mathbb{X}$, $\mathbb{L}$ and $\mathbb{Z}$. We use $\mathcal{F}(\mathcal{S})$ to denote all the finite subsets of a set $\mathcal{S}$ (including the empty set). On the other hand, the set exponential is defined as $\left[h\left(\cdot\right)\right]^{X}=\prod_{x\in X}h\left(x\right),$ and the inner product of two functions $f$ and $g$ is defined as $\langle f,g\rangle \triangleq \int f\left(x\right)g\left(x\right)dx.$ The Kronecker delta function with arbitrary argument is given by,
\begin{equation}
   \delta_{\mathcal{S}}\left(X\right)=\begin{cases} 1, & X=\mathcal{S}\\
0, & X\neq\mathcal{S} \end{cases},\end{equation}
and the set inclusion function is given by,
\begin{equation}1_{\mathcal{S}}\left(X\right)=\begin{cases} 1, & X\subseteq\mathcal{S}\\
0, & otherwise \end{cases}.
\end{equation}

A labeled single-object state can be written as $\bm{x}=\left(x,\ell\right)\in\mathbb{X}\times\mathbb{L}$. Conventionally, each label $\ell$ at time $k$ is an ordered pair $\ell=(t_{b},i)$, where $t_{b}\leq k$ denotes the time of birth and $i$ is an unique index to distinguish objects born at the same time. Let $\mathbb{B}$ denote the birth label space at current time step $k$, then the birth labels at time $k+1$ belong to the label space $\mathbb{B}_{+}=\left\{ \left(k+1,i\right):i\in\mathbb{N}\right\}$, and hence $\mathbb{L\cap B_{+}=\emptyset}$. The label space at time $k+1$ becomes $\mathbb{L_{+}=L\cup B_{+}}$. For compactness, we use subscript `+' to denote the next time step quantities.

The distinct label indicator \cite{Vo2013Labeled} is given as,
\begin{equation} \label{distinct label indicator} \Delta\left(\bm{X}\right)=\delta_{|\bm{X}|}\left(\left|\mathcal{L}\left(\bm{X}\right)\right|\right), \end{equation}
where $|\bm{X}|$ denotes the cardinality of a labeled set $\bm{X}$, and $\mathcal{L}:\mathbb{X\times L\rightarrow L}$ is a mapping from a labeled RFS to the labels, which satisfies the projection 
$\mathcal{L}(x,\ell)=\ell$. The distinct label indicator is used to ensure that $\bm{X}$ has distinct labels. 

The integral of a function $f:\mathcal{F}\left(\mathbb{X\times L}\right)\rightarrow\mathbb{R}$ is given by \cite{Vo2013Labeled}
    \begin{equation}
        \int f\left(\bm{X}\right)\delta \bm{X}=\sum_{i=0}^{\infty}\frac{1}{i!}\sum_{\left(\ell_{1},\ldots,\ell_{i}\right)\in\mathbb{L}^{i}}\int_{\mathbb{X}^{i}}f(\{ \left(x_{1},\ell_{1}\right), \ldots,\left(x_{i},\ell_{i}\right)\} )d\left(x_{1},\ldots,x_{i}\right).
   \label{eq:FISST integration} \end{equation}
    
In labeled multi-object Bayes filters, given the multi-object transition density $\bm{f}_{+}$, and the multi-object likelihood function $g_{+}$, a probability density function $\bm{\pi}$\footnote{This is not a probability density function but is equivalent to one as shown in \cite{Vo2005Sequential}. Hence, with a slight abuse of terminology, we regard this function as a probability density function.} on labeled multi-object state is propagated through time using the Bayes recursion \cite{Vo2013Labeled},
    \begin{equation} \bm{\pi}\left(\bm{X}_{+}\right) =\int\bm{f}_{+}\left(\bm{X}_{+}|\bm{X}\right)\bm{\pi}\left(\bm{X}\right)\delta\bm{X}, \label{eq:prior recursive} \end{equation}
    \begin{equation}\bm{\pi}_{+}\left(\bm{X}_{+}|Z_{+}\right) =\frac{g_{+}\left(Z_{+}|\bm{X}_{+}\right)\bm{\pi}\left(\bm{X}_{+}\right)}{\int g_{+}\left(Z_{+}|\bm{X}\right)\bm{\pi}\left(\bm{X}\right)\delta\bm{X}}, \label{eq:posterior recursive} \end{equation}
where the integrals in Eq. \eqref{eq:prior recursive} and Eq. \eqref{eq:posterior recursive} are the set integral defined in Eq. \eqref{eq:FISST integration}.

\subsection{The multi-object transition model}

Given a multi-object state $\bm{X}$ at the current time step, at the next time step, each object can either reappear with probability $p_{S}(x,\ell)$ and take on a new state $(x_{+},\ell_{+})$ computed via the single-object transition density $f_{S}(x_{+}|x,\ell)\delta_{\ell}(\ell_{+})$, or disappear from the sensor field of view with probability $q_{S}(x,\ell)=1-p_{S}(x,\ell)$. Overall, the multi-object transition density for a set of surviving objects, $\bm{X}_{S+}$,  can be written as,
    \begin{equation}\label{eq:survival distribution} 
               \bm{f}_{S+}\left(\bm{X}_{S+}|\bm{X}\right)=\Delta\left(\bm{X}_{S+}\right)\Delta\left(\bm{X}\right)1_{\mathcal{L}\left(\bm{X}\right)}\left(\mathcal{L}\left(\bm{X}_{S+}\right)\right)\left[\Phi_{S+}\left(\bm{X}_{S+}|\cdot\right)\right]^{\bm{X}},
       \end{equation}  
where,
    \begin{equation}
        \Phi_{S+}\left(\bm{X}_{S+}|x,\ell\right)= \sum_{\left(x_{+},\ell_{+}\right)\in\bm{X}_{S+}}\delta_{\ell}\left(\ell_{+}\right)p_{S}\left(x,\ell\right) f_{S+}\left(x_{+}|x,\ell\right)+\left[1-1_{\mathcal{L}\left(\bm{X}_{S+}\right)}\left(\ell\right)\right]q_{S}\left(x,\ell\right).\label{eq:Phi_S}
  \end{equation}
  
In addition to the surviving objects, new objects can also instantaneously appear in the field of view of the sensor at each time step. Let $\bm{X}_{B+}$ denote the labeled multi-object state representing those new-born objects, then its density is given as an LMB density of the form \cite{Vo2013Labeled}
    \begin{equation}
        \bm{f}_{B+}\left(\bm{X}_{B+}\right)=\Delta\left(\bm{X}_{B+}\right)\omega_{B}\left(\mathcal{L}\left(\bm{X}_{B+}\right)\right)\left[p_{B+}\right]^{\bm{X}_{B+}},
        \label{eq:birth distribution}
    \end{equation}
where
\begin{equation}
    \omega_{B}\left(L\right)=\left[1-r_{B+}\right]^{\mathbb{B}_{+}-L} 1_{\mathbb{B}+}\left(L\right) \left[r_{B+}\right]^{L},
\end{equation}
$r_{B+}(\ell)$ is the existence probability of a birth labeled $\ell$, and $p_{B+}(\cdot,\ell)$ is its unlabeled single-object spatial distribution.

Due to the independence between surviving objects and the new-born objects, multi-object transition density can be written as,
\begin{equation} \bm{f}_{+}\left(\bm{X}_{+}|\bm{X}\right)=\bm{f}_{S_{+}}\left(\bm{X}_{S+}|\bm{X}\right)\bm{f}_{B+}\left(\bm{X}_{B+}\right),\label{transition kernel without spawning} \end{equation}
where $\bm{X}_{S+}=\bm{X}_{+}\cap\left(\mathbb{X\times L}\right)$, and $\bm{X}_{B+}=\bm{X}_{+}\cap\left(\mathbb{X\times B_{+}}\right)$.

\subsection{ The multi-object measurement model}

Given a labeled set $\bm{X}$, each object $\bm{x}=(x,\ell)$ in this set can either generate a measurement $z\in\mathbb{Z}$ with a detection probability of $p_{D}(x,\ell)$ and likelihood of $g(z|x,\ell)$, or be miss-detected with the probability $q_{D}(x,\ell)=1-p_{D}(x,\ell)$. Further, sensor imperfections and environmental conditions also produce false measurements which are included in the observed measurement set. Hence, the likelihood of observing a set of measurements $Z$ given a labeled multi-object set $\bm{X}$ is given by \citep{Vo2013Labeled},
    \begin{equation} g\left(Z|\bm{X}\right)\propto\sum_{\theta\in\Theta\left(\mathcal{L}\left(\bm{X}\right)\right)}\prod_{\left(x,\ell\right)\in\bm{X}}\Psi_{Z}^{\left(\theta\left(\ell\right)\right)}\left(x,\ell\right), 
    \label{standard multitarget observation model} \end{equation}
    where:
    \begin{equation}
        \Psi_{Z}^{\left(\theta\left(\ell\right)\right)}\left(x,\ell\right)=\delta_{0}\left(\theta\left(\ell\right)\right)q_{D}\left(x,\ell\right)+\left(1-\delta_{0}\left(\theta\left(\ell\right)\right)\right) \frac{p_{D}\left(x,\ell\right)g\left(z_{\theta\left(\ell\right)}|x,\ell\right)}{\kappa\left(z_{\theta\left(\ell\right)}\right)};\label{eq:Psi_Z}
        \end{equation}
$\kappa(\cdot)$ is the Poisson intensity function; $\Theta$ is the set of all positive 1-1
association maps $\theta:\mathbb{L\rightarrow}\left\{ 0:|Z|\right\}$. \datrevise{The intensity function  is usually written as $\kappa=\lambda_{c}\mathcal{U}(\mathbb{Z})$, where $\lambda_{c}$ is the average clutter intensity and $\mathcal{U}(\mathbb{Z})$ denotes the uniform distribution on the measurement space.}

\section{An adaptive GLMB filter}
\begin{figure} \centering \includegraphics[width=4in]{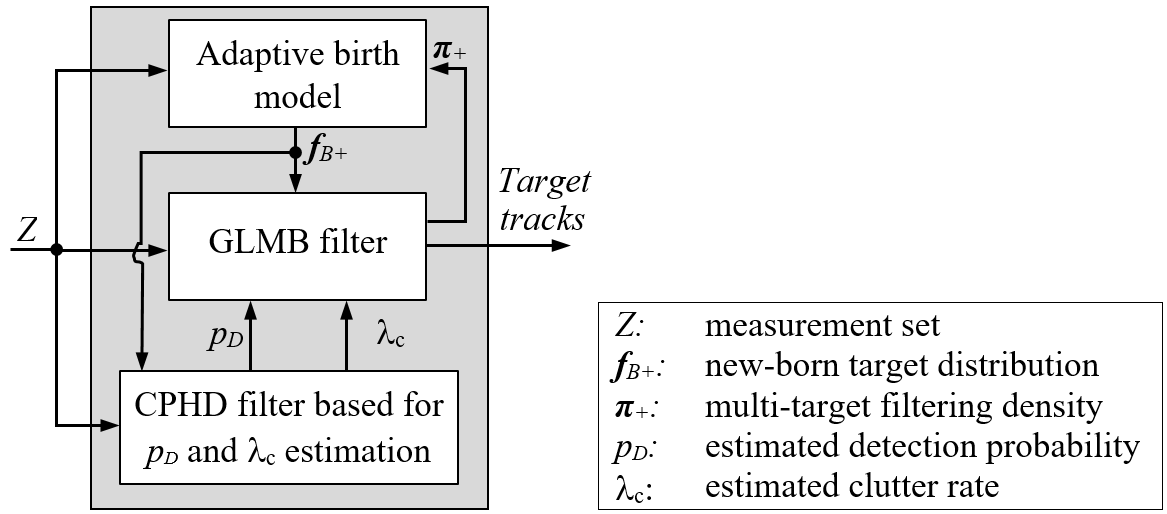} 
\caption{Block diagram of the proposed adaptive GLMB filter.}
\label{figure_2} \end{figure}

In this section, we provide
the detailed implementation of an adaptive GLMB filter that has the capability to track multiple objects with minimum prior knowledge from the users. In particular, we adopt a robust CPHD filter to estimate the clutter rate and average detection probability. This information is then bootstrapped into a standard GLMB filter to generate tracking results. As the prior knowledge of births is unknown, we utilize  
a measurement-driven birth model to  initialize new tracks. The structure of the proposed algorithm is shown in Fig. \ref{figure_2}. Thanks to the low complexity of the CPHD filter, the clutter rate, and detection probability are efficiently estimated on the fly while the GLMB filter allows the trajectories to be estimated given its labeled RFS formulation.

\subsection{GLMB filtering formulation}
\label{subsec:delta_GLMB} 
Let the multi-object density function at the current time step be written as a GLMB density \cite{Vo2013Labeled} i.e., 
\begin{equation}
    \bm{\pi}\left(\bm{X}\right)=\Delta\left(\bm{X}\right)\sum_{\left(I,\xi\right)\in\mathcal{F}\left(\mathbb{L}\right)\times\Xi}\omega^{\left(I,\xi\right)}\delta_{I}\left(\mathcal{L}\left(\bm{X}\right)\right)\left[p^{\left(\xi\right)}\right]^{\bm{X}},
    \label{eq: GLMB prior}
\end{equation}
where $I$ represents a set of labels, $\xi\in\Xi$ represents a history of association maps, and each $p^{\left(\xi\right)}(\cdot,\ell)$ represents a state distribution of an object on $\mathbb{X}\left(\text{with}\int p^{\left(\xi\right)}\left(x,\ell\right)dx=1\right)$. Each non-negative weight $\omega^{\left(I,\xi\right)}$ satisfies,
\begin{equation}
    \sum_{I\in\mathcal{F}\left(\mathbb{L}\right)}\sum_{\xi\in\Xi}\omega^{\left(I,\xi\right)}\left(L\right)=1. \label{standard prior weight}
\end{equation}

Then, by combining prediction and update steps in the Bayes recursion into a single step, the resultant measurement updated GLMB density can be written as \cite{Vo2017AnEfficient},
\begin{equation}
   \bm{\pi}_{+}\left(\bm{X}_{+}|Z_{+}\right)\propto\Delta\left(\bm{X}_{+}\right)\sum _{I,\xi,I_{+},\theta_{+}}\omega^{\left(I,\xi\right)}\omega_{Z_{+}}^{\left(I,\xi,I_{+},\theta_{+}\right)} 
   \delta_{I_{+}}\left[\mathcal{L}\left(\bm{X}_{+}\right)\right]\left[p_{Z_{+}}^{\left(\xi,\theta_{+}\right)}\right]^{\bm{X}_{+}},
 \label{GLMB_JPU_0} \end{equation}
where $I_{+}\in\mathcal{F}\left(\mathbb{L}_{+}\right), \theta_{+}\in\Theta_{+}\left(I_{+}\right)$ is a positive 1-1 association map $\theta_{+}:I_{+}\rightarrow\{0:|Z_{+}|\}$ with $Z_{+}$ is the observed measurement set at the next time step, and
\begin{align}\begin{split}
    \omega_{Z_{+}}^{\left(I,\xi,I_{+},\theta_{+}\right)} = & 1_{\Theta_{+}\left(I_{+}\right)}\left(\theta_{+}\right)\left[1-\bar{P}_{S}^{\left(\xi\right)}\right]^{I-I_{+}}\left[\bar{P}_{S}^{\left(\xi\right)}\right]^{I\cap I_{+}}\left[1-r_{B,+}\right]^{\mathbb{B}_{+}-I_{+}}\\& \times r_{B,+}^{\mathbb{B}_{+}\cap I_{+}}\left[\bar{\psi}_{Z_{+}}^{\left(\xi,\theta_{+}\right)}\right]^{I_{+}} \label{GLMB_JPU_1},
\end{split}     \\
    \bar{P}_{S}^{\left(\xi\right)}\left(\ell\right) =& \left\langle p^{\left(\xi\right)}\left(\cdot,\ell\right),p_{S}\left(\cdot,\ell\right)\right\rangle,\label{GLMB_JPU_2}\\
  \bar{\psi}_{Z_{+}}^{\left(\xi,\theta_{+}\right)}\left(\ell_{+}\right)= & \left\langle\bar{p}_{+}^{\left(\xi\right)}\left(\cdot,\ell_{+}\right),\psi_{Z_{+}}^{\left(\theta_{+}\left(\ell_{+}\right)\right)}\left(\cdot,\ell_{+}\right)\right\rangle,\label{GLMB_JPU_3}\\
  \bar{p}_{+}^{\left(\xi\right)}\left(x_{+},\ell_{+}\right)   =& 1_{\mathbb{B}_{+}}\left(\ell_{+}\right)p_{B,+}\left(x_{+},\ell_{+}\right)+1_{\mathbb{L}}\left(\ell_{+}\right)\frac{\left\langle p_{S}\left(\cdot,\ell_{+}\right)f_{+}\left(x_{+}|\cdot,\ell_{+}\right),p^{\left(\xi\right)}\left(\cdot,\ell_{+}\right)\right\rangle}{\bar{P}_{S}^{\left(\xi\right)}\left(\ell_{+}\right)}\label{GLMB_JPU_4}, \\   
  p_{Z_{+}}^{\left(\xi,\theta_{+}\right)}\left(x_{+},\ell_{+}\right) = & \frac{\bar{p}_{+}^{\left(\xi\right)}\left(x_{+},\ell_{+}\right)\psi_{Z_{+}}^{\left(\theta_{+}\left(\ell_{+}\right)\right)}\left(x_{+},\ell_{+}\right)}{\bar{\psi}_{Z_{+}}^{\left(\xi,\theta_{+}\right)}\left(\ell_{+}\right)}\label{GLMB_JPU_5}.
\end{align}

Given the GLMB filtering density, the estimated multi-object state and the trajectories can be extracted \cite{Vo2013Labeled,Nguyen2019GLMB}. We adopt the joint prediction and update approach with a Gibbs sampler to select significant components (ones with high weights) of the GLMB filtering density \cite{Vo2017AnEfficient}. Note that in standard GLMB filter formulation, the set of new births $\bm{X}_{B+}$ is assumed to be known to the filter. However, in our adaptive multi-object tracking approach, the birth objects are initiated from a measurement-driven birth model discussed in the subsection \ref{subsec:Meas-driv_mod}. Furthermore, the standard GLMB filter assumes that the detection probability $p_{D}$ and the average intensity of clutter $\lambda_{c}$ are also known. In the proposed adaptive GLMB filter, these parameters are bootstrapped from an independently run robust CPHD filter, as explained in subsection \ref{subsec:prob_dect_clutt}.

\subsection{ Measurement-driven birth model}\label{subsec:Meas-driv_mod} In this work, we adopt the measurement-driven (also called adaptive) birth model proposed in \cite{Reuter2014TheLabeled} to
initialize new tracks. The adaptive birth model is based on the intuition that the lower the probability $z\in Z$ is associated with an existing object at the current time step, the higher the probability that it is generated by a new birth at the next time step. Hence, each measurement in the current measurement set can be used to initialize a new track whose existence probability depends on the (track) association probability of the measurement. Specifically, given a GLMB density of the form in Eq. \ref{eq: GLMB prior}, the association probability \cite{Reuter2014TheLabeled}, $r_{U}(z)$, of a measurement $z$ is given by,
    \begin{equation}
        r_{U}(z)=\sum_{I\in\mathcal{F}(\mathbb{L})} \sum_{\theta\in\Xi_{I,k}}1_{\theta}(z)\omega^{(I,\theta)},    
        \label{association probability}
    \end{equation}
where $\Xi_{I,k}$ is the set of association maps for a set of labels $I$ at the current time step. The inclusion function $1_{\theta}(z)$ ensures that the sum of weights only considers those hypotheses that assign the measurement $z$ to one of its tracks. Then, the probability of birth of an object based on the measurement $z$ at the next time step, is given by \cite{Reuter2014TheLabeled},
    \begin{equation}
        r_{B+}(z)=\min\left(r_{B,\textrm{max}},\frac{1-r_{U}(z)}{\sum_{\xi\in Z}1-r_{U}(\xi)} \cdot \lambda_{B+}\right),
        \label{existing probability}
    \end{equation}
where $\lambda_{B+}$ is the expected number of birth objects at the next time step, and $r_{B,\textrm{max}}\in[0,1]$ is the maximum birth probability. Further, the intensity of this birth LMB is also supplied to the robust CPHD filter presented in the next subsection.

\subsection{Detection probability and clutter rate estimation}\label{subsec:prob_dect_clutt} 
The CPHD filter tracks multiple objects assuming that the elements of the multi-object RFS are identically and independently distributed (i.i.d.), with an arbitrary cardinality distribution (an instantiation of an i.i.d. cluster process). The CPHD filter is computationally efficient as it alleviates the tracks to measurement association problem, despite not being able to provide track identities. Therefore, we adopt the robust CPHD \cite{Mahler2011CPHD} for estimating the clutter rate and the detection probability in the proposed adaptive GLMB filter.


The closed-form filtering solution of this filter is obtained on the augmented hybrid state space consisting of the actual object state space and clutter object state space. Specifically, let the state space of actual objects, clutter generated objects and the detection probability be respectively denoted as $\mathbb{X}^{(1)},\mathbb{X}^{(0)}$ and $\mathbb{X}^{(\Delta)}=[0,1]$. Then, the hybrid state space is given by \cite{Mahler2011CPHD}, \begin{align} \mathbb{X}^{\left(h\right)}=\left(\mathbb{X}^{\left(1\right)}\times\mathbb{X}^{\left(\Delta\right)}\right)\uplus\left(\mathbb{X}^{\left(0\right)}\times\mathbb{X}^{\left(\Delta\right)}\right). \label{hybrid and augmented state space} \end{align}

Note that the behaviors of actual objects and the clutter objects are assumed to be statistically independent. Then, the CPHD filter is adopted to estimate the PHD and the cardinality distribution of the multi-object filtering density on the hybrid state space. Therefore, cardinality distributions of clutter and actual objects can be  individually extracted, from which the clutter rate can be inferred.

To facilitate the discussion on the recursive computation of the PHD and cardinality distributions for the robust CPHD filter, we define the following dynamic and measurement models on the augmented state space. A single actual object state is defined as $x_{r}^{(h)}=(x_{r},a)\in\mathbb{X}^{\left(1\right)}\times\mathbb{X}^{\left(\Delta\right)}$, where $a\in\mathbb{X}^{(\Delta)}$ is its detection probability and a single clutter object state is defined as $x_{c}^{(h)}=(x_{c},b)\in\mathbb{X}^{\left(0\right)}\times\mathbb{X}^{\left(\Delta\right)}$, where $b\in\mathbb{X}^{(\Delta)}$ is its detection probability. Due to the statistical independence (between actual and clutter objects), the integral of a function $f^{\left(h\right)}:\mathbb{X}^{\left(h\right)}\rightarrow\mathbb{R}$ can be written as \cite{Mahler2011CPHD},
\begin{equation}
          \int_{\mathbb{X}^{\left(h\right)}}f^{\left(h\right)}\left(x^{\left(h\right)}\right)dx^{\left(h\right)}=\int_{\mathbb{X}^{\left(\Delta\right)}}\int_{\mathbb{X}^{\left(1\right)}}f^{\left(h\right)}\left(x_{r},a\right)dx_{r}da+\int_{\mathbb{X}^{\left(\Delta\right)}}\int_{\mathbb{X}^{\left(0\right)}}f^{\left(h\right)}\left(x_{c},b\right)dx_{c}db.    \label{hybrid integral function}
 \end{equation}
 
The joint single object survival probability is defined as follows \cite{Mahler2011CPHD},
    \begin{equation}
        p_{S+}^{\left(h\right)}(x^{(h)}) =\begin{cases} p_{S+}^{\left(1\right)}, & \text{if}\quad  x^{\left(h\right)}\in\mathbb{X}^{\left(1\right)}\times\mathbb{X}^{\left(\Delta\right)}\\ p_{S+}^{\left(0\right)}, & \text{if}\quad   x^{\left(h\right)}\in\mathbb{X}^{\left(0\right)}\times\mathbb{X}^{\left(\Delta\right)} \end{cases}.
    \end{equation}
    
The joint transition density is defined as,

\begin{equation}
   f_{+}^{\left(h\right)}\left(x_{+}^{\left(h\right)}|x^{\left(h\right)}\right)=\begin{cases}
    f_{+}^{\left(1\right)}\left(x_{r+}|x_{r}\right)f_{+}\left(a_{+}|a\right), & \text{if }\quad\begin{aligned}x_{+}^{\left(h\right)} & =\left(x_{r+},a_{+}\right)\text{and}\\
    x^{\left(h\right)} & =\left(x_{r},a\right)\in\mathbb{X}^{\left(1\right)}\times\mathbb{X}^{\left(\Delta\right)}
    \end{aligned}
    \\
    f_{+}^{\left(0\right)}\left(x_{c+}|x_{c}\right), & \text{if }\quad\begin{aligned}x_{+}^{\left(h\right)} & =\left(x_{c+},b_{+}\right)\text{and}\\
    x^{\left(h\right)} & =\left(x_{c},b\right)\in\mathbb{X}^{\left(0\right)}\times\mathbb{X}^{\left(\Delta\right)}
    \end{aligned}
    \\
    0, & \begin{aligned}\textrm{otherwise}\end{aligned}
    \end{cases}.
    \label{Joint trans dens}
\end{equation}

The joint birth intensity at the next time step is defined as,
\begin{equation}
    \begin{split}
       \gamma_{+}^{\left(h\right)}\left(x_{+}^{\left(h\right)}\right) =\begin{cases} \gamma_{+}^{\left(1\right)}\left(x_{r+},a_+\right), &\text{if}\quad x_{+}^{\left(h\right)}=\left(x_{r+},a_+\right)\in\mathbb{X}_{+}^{\left(1\right)}\times\mathbb{X}_{+}^{\left(\Delta\right)} \\ \gamma_{+}^{\left(0\right)}\left(x_{c+},b\right), &\text{if}\quad x_{+}^{\left(h\right)}=\left(x_{c+},b\right)\in\mathbb{X}_{+}^{\left(0\right)}\times\mathbb{X}_{+}^{\left(\Delta\right)} \end{cases},
       \label{joint birth intensity}     
    \end{split}
\end{equation} 
and the joint birth cardinality distribution is defined by a convolution as follows,
\begin{align}\rho_{X_{B+}}^{\left(h\right)}\left(n^{\left(h\right)}\right) &=\left(\rho_{X_{B+}}^{\left(1\right)}*\rho_{X_{B+}}^{\left(0\right)}\right)\left(n^{\left(h\right)}\right).\label{joint birth cardinality} \end{align}

The joint detection probability is defined as,
\begin{equation}
    p_{D}^{\left(h\right)}\left(x^{\left(h\right)}\right) =\begin{cases} a, &\text{if}\quad  x^{\left(h\right)}=\left(x_{r},a\right)\in\mathbb{X}^{\left(1\right)}\times\mathbb{X}^{\left(\Delta\right)}\\
    b,  &\text{if}\quad x^{\left(h\right)}=\left(x_{c},b\right)\in\mathbb{X}^{\left(0\right)}\times\mathbb{X}^{\left(\Delta\right)} \end{cases},
    \label{joint detection probability}
\end{equation} 
and the joint measurement likelihood function is defined as,
\begin{align}
    g^{\left(h\right)}\left(z|x^{\left(h\right)}\right) &=\begin{cases} g^{\left(1\right)}\left(z|x^{\left(h\right)}\right), &\text{if}\quad x^{\left(h\right)}=\left(x_{r},a\right)\in\mathbb{X}^{\left(1\right)}\times\mathbb{X}^{\left(\Delta\right)}\\
    g^{\left(0\right)}\left(z\right), &\text{if}\quad x^{\left(h\right)}=\left(x_{c},b\right)\in\mathbb{X}^{\left(0\right)}\times\mathbb{X}^{\left(\Delta\right)}\label{joint likelihood function} \end{cases}.
\end{align}
Given the defined transition densities, the PHDs at current time step ($\nu^{(0)}$,  $\nu^{(1)}$), and the cardinality distributions ($\rho^{(0)}$, $\rho^{(1)}$), the predicted PHDs and cardinality distributions can be computed as \cite{Mahler2011CPHD},
\begin{align}\begin{split}
  \nu_{+}^{\left(1\right)}\left(x_{r+},a_{+}\right) =&\int\int_{0}^{1}p_{S+}^{\left(1\right)}\left(x_{r}\right)f_{+}^{\left(\Delta\right)}\left(a_{+}|a\right)f_{+}^{\left(1\right)}\left(x_{r+}|x_{r}\right)\nu^{\left(1\right)}\left(a,x_{r}\right)dadx_{r}\\ & +\gamma_{+}^{\left(1\right)}\left(x_{r+},a_{+}\right),
\end{split}
\\
\nu^{\left(0\right)}_{+}\left(b\right)  =&\gamma_{+}^{(0)}(b)+p_{S+}^{(0)}\nu^{(0)}\left(b\right),\\
\rho_{+}^{\left(h\right)}\left(n^{\left(h\right)}\right)  =&\sum_{j=0}^{n^{(h)}}\rho_{X_{B+}}^{\left(h\right)}\left(n^{\left(h\right)}-j\right)\sum_{\ell=j}^{\infty}\mathcal{C}_{j}^{\ell}\rho^{\left(h\right)}\left(\ell\right)\left(1-\phi\right)^{\ell-j}\phi^{j},  
\end{align}
where $C^{\ell}_{j}=\frac{\ell!}{j!(\ell-j)!}$ is the binomial coefficient, and
\begin{align} \phi=\left(\frac{\left\langle\nu^{\left(1\right)},p_{S+}^{\left(1\right)}\right\rangle+\left\langle\nu^{\left(0\right)},p_{S+}^{\left(0\right)}\right\rangle}{\left\langle1,\nu^{\left(1\right)}\right\rangle+\left\langle1,\nu^{\left(0\right)}\right\rangle}\right). \end{align}
Then, given the measurements set $Z_{+}$, the measurement updated PHDs and cardinality distributions can be computed via \cite{Mahler2011CPHD},
 \begin{align}\begin{split}
    \nu_{+}^{\left(1\right)}\left(x_{r+},a_{+}|Z_{+}\right)  = &\nu_{+}^{\left(1\right)}\left(x_{r+},a_{+}\right)\left[\frac{\left(1-a_{+}\right)\frac{\left\langle\Gamma_{+}^{\left(1\right)}\left[\nu_{+}^{\left(h\right)},Z_{+}\right],\rho_{+}^{\left(h\right)}\right\rangle}{\left\langle\Gamma_{+}^{\left(0\right)}\left[\nu_{+}^{\left(h\right)},Z_{+}\right],\rho_{+}^{\left(h\right)}\right\rangle}}{\left\langle1,\nu_{+}^{\left(1\right)}\right\rangle+\left\langle1,\nu_{+}^{\left(0\right)}\right\rangle}\right. \\&\left. + \sum_{z\in Z_{+}}\frac{a_{+}\cdot g_{+}\left(z|x_{r+}\right)}{\left\langle\nu_{+}^{\left(0\right)},p_{D+}^{\left(0\right)}g_{+}^{\left(0\right)}\right\rangle+\left\langle\nu_{+}^{\left(1\right)},p_{D+}^{\left(1\right)}g_{+}^{\left(1\right)}\left(z|\cdot\right)\right\rangle}\right],  
 \end{split} \\
 \begin{split}
     \nu_{+}^{\left(0\right)}\left(b_{+}|Z_{+}\right)  =&\nu_{+}^{(0)}\left(b_{+}\right)\left[\frac{\left(1-b_{+}\right)\frac{\left\langle\Gamma_{+}^{\left(1\right)}\left[\nu_{+}^{\left(h\right)},Z_{+}\right],\rho_{+}^{\left(h\right)}\right\rangle}{\left\langle\Gamma_{+}^{\left(0\right)}\left[\nu_{+}^{\left(h\right)},Z_{+}\right],\rho_{+}^{\left(h\right)}\right\rangle}}{\left\langle1,\nu_{+}^{\left(1\right)}\right\rangle+\left\langle1,\nu_{+}^{\left(0\right)}\right\rangle}\right.\\&\left.+\sum_{z\in Z_{+}}\frac{b_{+}\cdot g_{+}^{\left(0\right)}\left(z\right)}{\left\langle\nu_{+}^{\left(0\right)},p_{D+}^{\left(0\right)}g_{+}^{\left(0\right)}\right\rangle+\left\langle\nu_{+}^{\left(1\right)},p_{D+}^{\left(1\right)}g_{+}^{\left(1\right)}\left(z|\cdot\right)\right\rangle}\right], 
 \end{split} \end{align}
 \begin{align}
     \rho_{+}^{\left(h\right)}\left(n^{\left(h\right)}\right) & =\begin{cases} 0, &\text{if}\quad n^{\left(h\right)}<|Z_{+}|,\\ \frac{\rho_{+}^{\left(h\right)}\left(n^{\left(h\right)}\right)\Gamma_{+}^{\left(0\right)}\left[\nu_{+}^{\left(h\right)},Z_{+}\right]\left(n^{\left(h\right)}\right)}{\left\langle\rho_{+}^{\left(h\right)},\Gamma_{+}^{\left(0\right)}\right\rangle}, &\text{if}\quad n^{\left(h\right)}\geq|Z_{+}|, \end{cases}
 \end{align}
where\diluka{,}
\begin{equation}
  \Gamma_{+}^{\left(s\right)}\left[\nu_{+}^{\left(h\right)}Z_{+}\right]\left(n^{\left(h\right)}\right) =\begin{aligned}\begin{cases}
    0, &\text{if}\quad n^{\left(h\right)}<|Z_{+}|+s \\ 
P_{|Z_{+}|+s}^{\left(n^{\left(h\right)}\right)}\Phi_{+}^{\left(n^{\left(h\right)}-\left(|Z_{+}|+s\right)\right)},&\text{if}\quad  n^{\left(h\right)}\geq|Z_{+}|+s 
  \end{cases}, \end{aligned}
\end{equation} 
\begin{align}
\Phi &= 1-\frac{\left\langle\nu_{+}^{\left(1\right)},p_{D+}^{\left(1\right)}\right\rangle+\left\langle\nu_{+}^{\left(0\right)},p_{D+}^{\left(0\right)}\right\rangle}{\left\langle1,\nu_{+}^{\left(1\right)}\right\rangle+\left\langle1,\nu_{+}^{\left(0\right)}\right\rangle},
\end{align}
and $P^{n}_{j}=\frac{n!}{(n-j)!}$ is the permutation coefficient.

\diluka{At each time step, the clutter rate can be inferred via the updated cardinality distribution (from which the average clutter intensity $\lambda_{c}$ can be computed). In particular, the average number of \diluka{clutter objects at any given time} can be computed using \cite{Mahler2011CPHD},
\begin{equation}
    \bar{N}_{+}^{(0)}=\left\langle\nu_{+}^{\left(0\right)},p_{D+}^{\left(0\right)}\right\rangle.
\end{equation} Furthermore, the average detection probability $\bar{p}_{D+}^{(1)}$ can be computed by averaging the estimated detection probability of actual objects.}


\section{Numerical study}\label{sec:Numerical-study}

In this section, we demonstrate the performance of the proposed adaptive GLMB filter in multi-object tracking scenarios with linear and non-linear object dynamics, and compare the performance against ideal-GLMB and ideal-CPHD filters (parameters are known to these filters).

\subsection{Linear dynamic model} 
\begin{figure} \centering
    \includegraphics[width=0.5\textwidth]{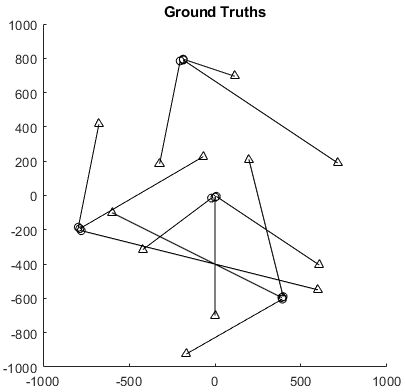} \caption{Ground true tracks in linear dynamic scenario ($\bigcirc:$ the track's initial position, $\bigtriangleup:$ the track's ending position). }
    \label{figure_3}
\end{figure}

A multi-object tracking scenario involving 
12 objects having linear motion is investigated in this experiment. The surveillance region is an area of $[-1000,1000]m\times[-1000,1000]m$, and a total of 100 time steps are simulated with sampling interval of $\Delta=1s$. The ground true tracks and their starting/ending positions are shown in Fig. \ref{figure_3}. The simulated value of the clutter rate is set to 50 false alarms per scan and the simulated detection probability is set to $0.95$.

The single-object state is a 4-D vector consisting of planar position and velocity of the object, i.e. $x=[p_{h},p_{v},\dot{p}_{h},\dot{p}_{v}]^{T}$, where $p_{h}$, $p_{v}$ are the position of the object in horizontal and vertical coordinates and $\dot{p}_{h}$, $\dot{p}_{v}$ are their corresponding velocities, and  $T$ denotes matrix transpose operation. The single-object transition density of the kinematic state is given as a Gaussian distribution, i.e.
\begin{align}
 f_{+}\left(x_{+}|x\right) =&\mathcal{N}\left(x_{+};Fx,Q\right)
\end{align}
where $F=\left[\begin{array}{cc} I_{2} & \Delta I_{2}\\ 0_{2} & I_{2} \end{array}\right]$, $Q=\sigma_{\nu}^{2}\left[\begin{array}{cc} \frac{\Delta^{4}}{4}I_{2} & \frac{\Delta^{3}}{2}I_{2}\\ \frac{\Delta^{3}}{2}I_{2} & \Delta^{2}I_{2} \end{array}\right]$
 with $\sigma_{\nu}=5m/s$, $I_n$ is the $n$-D identity matrix, and \datrevise{$\mathcal{N}\left(\cdot;m,P\right)$ denotes a Gaussian distribution with mean $m$ and covvariance matrix $P$.} In the proposed GLMB filter, the detection probability of each object is modeled by a beta distribution with the initial parameters of $s=90$ and $t=10$ (yielding an initial mean value of 0.9). The probability of survival $p_S$ is set to $0.99$. The detection probability and survival probability of clutter objects are set to $0.95$ and $0.9$ respectively. The maximum birth probability $r_{B,\textrm{max}}$, is set to 0.01, and the birth covariance matrix of a new-born 
object is set to $P_{B}=\text{diag}([10,10,10,10])$, where `diag' converts a vector to a diagonal matrix.

A measurement $z=[z_{h},z_{v}]^{T}$ is the observed planar position of an object. The the single-object measurement likelihood model is given as follows,
\begin{align}
g\left(z|x\right) =&\mathcal{N}\left(z;Hx,R\right),    
\end{align}
where $H=\left[\begin{array}{cc} I_{2} & 0_{2}\end{array}\right]$ and $R=\sigma_{\varepsilon}^{2}I_{2}$ with $\sigma_{\varepsilon}=15m$.

\begin{figure}
    \centering
    \begin{subfigure}{0.47\textwidth}\centering
        \includegraphics[width=\textwidth]{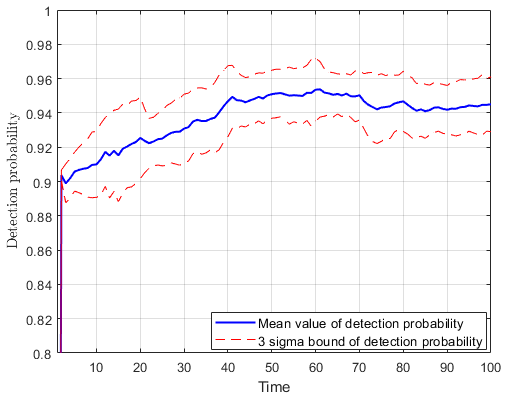} 
    \label{est PD linear}
    \vspace{-0.7cm}\caption{}
    \end{subfigure}
    \hspace{0.7cm}
    \begin{subfigure}{0.47\textwidth}\centering
    \includegraphics[width=\textwidth]{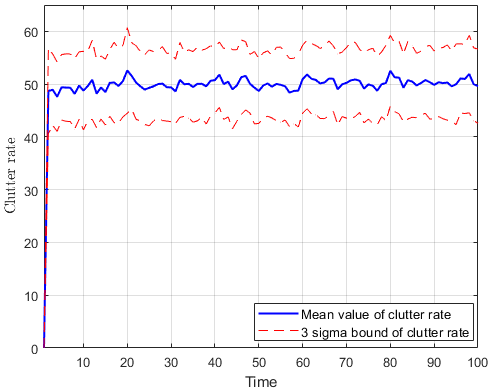}
    \label{est lambda linear}
    \vspace{-0.7cm}\caption{}
    \end{subfigure}
    \caption{Estimated detection probability and clutter rate from DP-GLMB filter in linear dynamic scenario.}
    \label{estimate pD and lambda}
\end{figure}

The performance of the proposed GLMB filter (referred to as DP-GLMB from here on, where DP stands for the dynamic parameter) is compared with the ideal CPHD and GLMB filters (these filters are supplied with correct simulated values of birth locations, detection probability and clutter rate). The maximum numbers of components for both DP-GLMB and ideal-GLMB are set to $5000$. The experiment is conducted over 100 Monte Carlo runs. The estimated probability of detection from the DP-GLMB filter is shown in Fig. \ref{estimate pD and lambda}(a) and its estimated clutter rate  is shown in Fig. \ref{estimate pD and lambda}(b). \diluka{The estimation error between the ground truth and the estimated multi-object state is captured using both optimal sub-pattern assignment (OSPA) \cite{Schuhmacher2008AConsistent} and OSPA$^{(2)}$ \cite{Beard2020ASolution} metrics. Note that OSPA$^{(2)}$ error can only be computed for tracking results that contains object identities.}

\begin{figure}
    \centering
    \begin{subfigure}{0.47\textwidth} \centering \includegraphics[width=\textwidth]{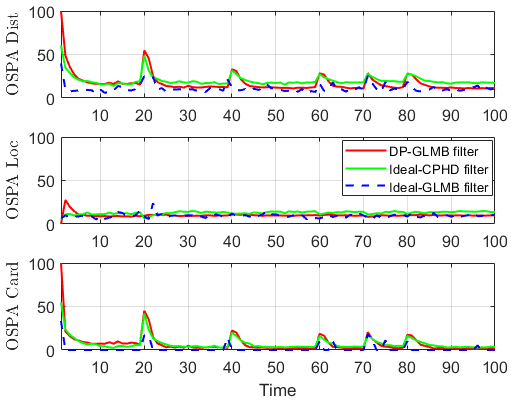}     \label{OSPA linear}
    \vspace{-0.7cm}\caption{}
    \end{subfigure}
    \hspace{0.7cm}
    \begin{subfigure}{0.47\textwidth} \centering \includegraphics[width=\textwidth]{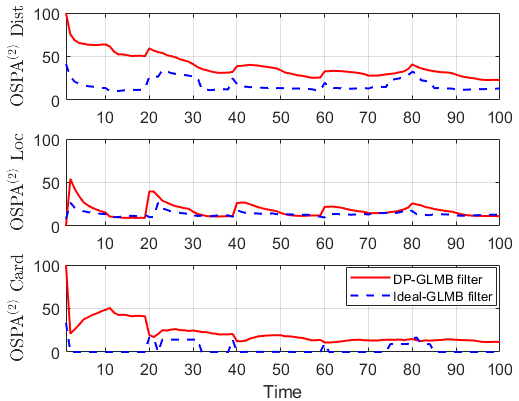}    \label{OSPA2 linear}
    \vspace{-0.7cm}\caption{}
    \end{subfigure}
       \caption{Mean OSPA and OSPA$^{(2)}$ errors of different filters in linear dynamic scenario. Results from ideal-CPHD filter are excluded in OSPA$^{(2)}$ error computation as CPHD filter does not provide labeled tracks.}
    \label{OSPA and OSPA2} 
\end{figure}

\begin{figure}
    \centering \includegraphics[width=0.5\textwidth]{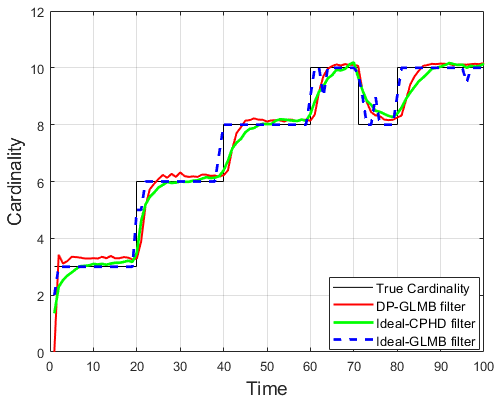} \caption{Mean estimated cardinality from different filters in linear dynamic scenario.}
    \label{ekf cardinality}
\end{figure}

Both OSPA and OSPA$^{(2)}$ metric settings have the norm order of $1$ and cut-off of $100$. The window-length for OSPA$^{(2)}$ metric is set to 10 time steps. Fig. \ref{OSPA and OSPA2}(a) shows the mean OSPA errors of 3 filters over 100 Monte Carlo runs while Fig. \ref{OSPA and OSPA2}(b)  shows their OSPA$^{(2)}$ errors.  Finally, the estimated cardinality over time from the filters are shown in Fig. \ref{ekf cardinality}.

\subsection{Non-linear dynamic model}
\begin{figure} \centering \includegraphics[width=0.6\textwidth]{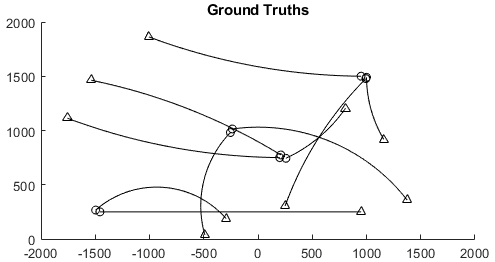} \caption{Ground true tracks in non-linear dynamic scenario ($\bigcirc:$ the track's initial position, $\bigtriangleup:$ the track's ending position).} \label{ukf groundtruth} \end{figure}

In this experiment, a multi-object tracking scenario involving 10 objects with each having a constant turn motion model is investigated. The surveillance area is a half-disk with a radius of $2000m$. The ground true tracks, their starting and ending positions are shown in Fig. \ref{ukf groundtruth}. The values  of the detection probability and clutter rate used to simulate the scenario are set to $0.95$ and $50$ false alarms per scan, respectively. The simulated tracking duration is also 100$s$ with sampling interval of $\Delta=1s$.

Single-object state is represented by a 5-D vector, i.e.
$x=[p_{h},p_{v},\dot{p}_{h},\dot{p}_{v},\omega]^{T}$, where $\omega$ is the turn rate of the object. The single-object transition model is given as,
\begin{align}
    f_{+}\left(x_{+}|x\right)=&\thinspace\mathcal{N}\left(x_{+};F\left(x,\omega\right)x,Q\right),
\end{align}
where 
\begin{equation*}
    F\left(x,\omega\right)=\left[\begin{array}{ccccc} 1 & \frac{sin\left(\Delta\omega\right)}{\omega} & 0 & -\frac{1-cos\left(\Delta\omega\right)}{\omega} & 0\\ 0 & cos\left(\Delta\omega\right) & 0 & -sin\left(\Delta\omega\right) & 0\\ 0 & \frac{1-cos\left(\Delta\omega\right)}{\omega} & 1 & \frac{sin\left(\Delta\omega\right)}{\omega} & 0\\ 0 & sin\left(\Delta\omega\right) & 0 & cos\left(\Delta\omega\right) & 0\\ 0 & 0 & 0 & 0 & 1 \end{array}\right], \thinspace
    Q=\left[\begin{array}{cc} \sigma_{\omega}^{2}GG^{T} & 0\\ 0 & \sigma_{v}^{2} \end{array}\right],
    G=\left[\begin{array}{cc} \Delta^{2}/2 & 0\\ \Delta & 0\\ 0 & \Delta^{2}/2\\ 0 & \Delta \end{array}\right],
\end{equation*}
$\sigma_{\nu}=5 m/s$, and $\sigma_{\omega}=\pi/180$ $rad/s$. The detection probability of each object in our filter is also modeled by a beta distribution with the initial parameters as in previous experiment. The survival probability $p_S$ is set to $0.99$.  The probability of birth is capped at $r_{B,\textrm{max}}= 0.02$, and the covariance matrix of the birth kinematic state is set to $P_{B}=\text{diag}([50,50,50,50,\pi/30])$.

A measurement is given by $z=\left[\theta,r\right]^{T}$, where $\theta$ is the observed bearing angle and $r$ is the observed range. Specifically, the non-linear measurement likelihood model is given as
\begin{align}
    g\left(z|x\right) = &\thinspace \mathcal{N}\left(z;\mu(x),R\right),  
\end{align}
where $\mu(x)=[\text{arctan}(p_{h}/p_{v}),\sqrt{p_{h}^{2}+p_{v}^{2}}]$, $R=\text{diag}([\sigma_{\theta}^2,\sigma_{r}^2])$ with $\sigma_{\theta}=\pi/180$ and $\sigma_{r}=5m$.

\begin{figure}[ht]
    \centering
    \begin{subfigure}{0.47\textwidth} \centering \includegraphics[width=\textwidth]{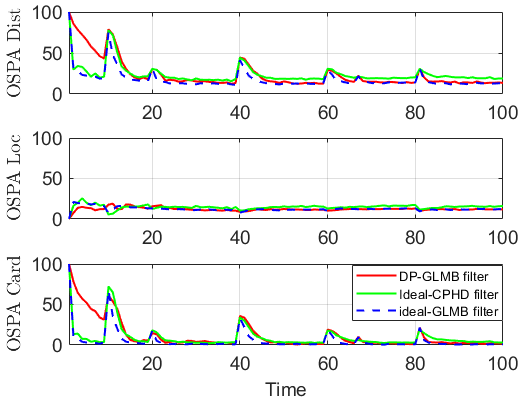}  \label{OSPA nonlinear}
    \vspace{-0.7cm}\caption{}
    \end{subfigure}
    \hspace{0.7cm}
    \begin{subfigure}{0.47\textwidth} \centering \includegraphics[width=\textwidth]{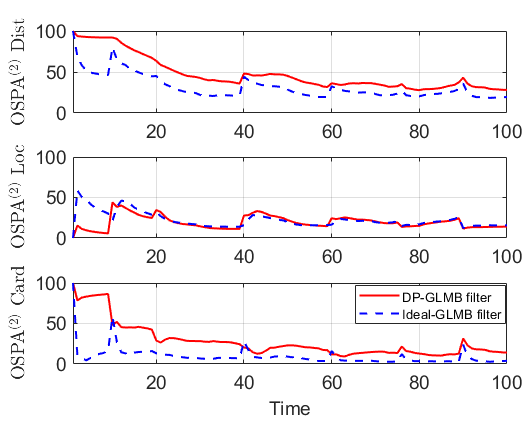}  \label{OSPA2 nonlinear}
    \vspace{-0.7cm}\caption{}
    \end{subfigure}
     \caption{ Mean OSPA and OSPA$^{(2)}$ errors of different filters in the non-linear dynamic scenario. Results from ideal-CPHD filter are excluded in OSPA$^{(2)}$ error computation as CPHD filter does not provide labeled tracks.}
   \label{ukf-OSPA-and-OSPA2}
   \end{figure}

\begin{figure}[ht]
    \centering \includegraphics[width=0.5\textwidth]{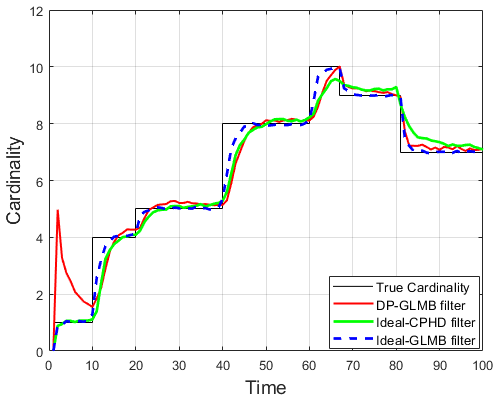} \caption{Mean estimated cardinality from different filters in non-linear dynamic scenario.} \label{ukf cardinality} 
\end{figure}

The performance of the DP-GLMB filter is also compared with the ideal-GLMB and ideal-CPHD filters. The maximum numbers of components for DP-GLMB and ideal-GLMB filters are both set to $5000$. The mean OSPA and OSPA$^{(2)}$ errors over $100$ Monte Carlo runs for different filters are shown in Figs. \ref{ukf-OSPA-and-OSPA2}(a) and \ref{ukf-OSPA-and-OSPA2}(b), respectively. The cardinality estimates from different filters over time are shown in Fig. \ref{ukf cardinality}, while the estimated probability of detection and clutter rate are shown in Figs. \ref{ukf estimate pD and Lambda}(a) and \ref{ukf estimate pD and Lambda}(b).

\subsection{Discussion on simulation results} 
\diluka{Observing all the simulation results, it is clear that estimates from both DP-GLMB and ideal-GLMB filters exhibit lower OSPA errors than that of ideal-CPHD filter (Figs. \ref{OSPA and OSPA2}  and \ref{ukf-OSPA-and-OSPA2}). The OSPA errors for DP-GLMB and ideal-GLMB filters are almost identical for both tracking scenarios. However, DP-GLMB exhibits higher error at the first few time steps due to the initial uncertainty measurement-driven birth model.}

\diluka{Both the ideal-GLMB filter and DP-GLMB filter produce almost identical OSPA$^{(2)}$ localization errors. The ideal-GLMB filter produces lower OSPA$^{(2)}$ cardinality errors, primarily due to the delay in track initialization and termination processes in the DP-GLMB filter (Figs. \ref{ekf cardinality} and \ref{ukf cardinality}). Therefore, the ideal-GLMB filter produces slightly lower overall OSPA$^{(2)}$ errors than the DP-GLMB filter. The difference in performance can be attributed to the fact that the DP-GLMB filter incurs higher uncertainty than the ideal-GLMB filter due to unknown detection probability, clutter rate and birth locations. However, note that the OSPA$^{(2)}$ errors in the DP-GLMB filter decrease over time as more and more data is processed, resolving the high initial uncertainty (in detection, clutter rate and birth locations).} 

\begin{figure}[ht]
    \centering
    \begin{subfigure}{0.47\textwidth}\centering \includegraphics[width=\textwidth]{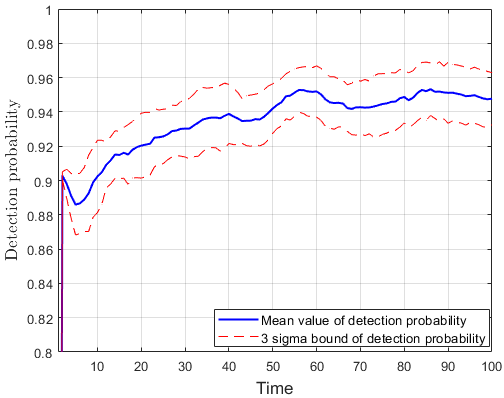}   \label{est pD nonlinear}
    \vspace{-0.7cm}\caption{}
    \end{subfigure}
    \hspace{0.7cm}
    \begin{subfigure}{0.47\textwidth}\centering \includegraphics[width=\textwidth]{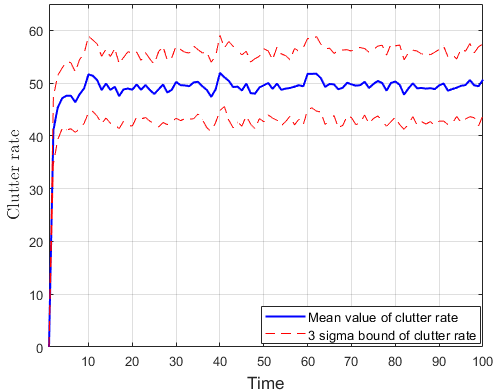}   \label{est lambda nonlinear}
    \vspace{-0.7cm}\caption{}
    \end{subfigure}
    \caption{Estimated detection probability and clutter rate from DP-GLMB filter in non-linear dynamic scenario.}
   \label{ukf estimate pD and Lambda} 
\end{figure}

\diluka{Furthermore, the estimated detection probability parameter of the DP-GLMB filter reaches its actual value of 0.95 from the initial value of 0.9 (Figs. \ref{estimate pD and lambda}(a) and \ref{ukf estimate pD and Lambda}(a)). The clutter rate parameter also reaches the true mean value of 50 clutter points per scan for both tracking scenarios (Figs. \ref{estimate pD and lambda}(b) and \ref{ukf estimate pD and Lambda}(b)).
}

\section{Conclusion}\label{sec:Conclusions} 
\diluka{This paper proposes an online adaptive multi-object tracking filter based on the GLMB filter to accommodate unknown detection probability, clutter rate, and birth locations. Compared to a standard GLMB filter, our approach mitigates the need for  off-line training or tediously fine-tuning (of) parameters by utilizing a robust CPHD filter and a measurement-driven birth model. The experimental results showed that the proposed filter exhibits comparable performances to an ideal GLMB filter (i.e., a standard GLMB filter supplied with correct parameters) and performs better than an ideal CPHD filter. Future avenues of research with this filter include catering for multi-sensor multi-object tracking, object spawning and multi-scan state estimation.}

 \bibliographystyle{IEEEtran}
 \bibliography{refbib111}

\end{document}